\documentclass[a4paper,11pt]{article}
\pdfoutput=1 

\usepackage{jinstpub} 
\graphicspath{ {figures/} }
\usepackage[hang,tight]{subfigure}
\usepackage[normalem]{ulem}
\usepackage{lineno}

\usepackage{placeins} 

\begin{document}


\title{\boldmath Lithium-loaded Liquid Scintillator Production for the PROSPECT experiment}

\collaboration{The PROSPECT Collaboration}
\author[p]{J.\,Ashenfelter,}
\author[m]{A.\,B.\,Balantekin,}
\author[p]{H.\,R.\,Band,}
\author[f]{C.\,D.\,Bass,}
\author[g]{D.\,E.\,Bergeron,}
\author[j]{D.\,Berish,}
\author[a]{L.\,J.\,Bignell,}
\author[e]{N.\,S.\,Bowden,}
\author[e]{J.\,P.\,Brodsky,}
\author[h]{C.\,D.\,Bryan,}
\author[a]{C.\,Camilo Reyes,}
\author[a]{S.\,Campos,}
\author[n]{J.\,J.\,Cherwinka,}
\author[e]{T.\,Classen,}
\author[c]{A.\,J.\,Conant,}
\author[o]{D.\,Davee,}
\author[i]{D.\,Dean,}
\author[h]{G.\,Deichert,}
\author[a]{R.\,Diaz Perez,}
\author[a]{M.\,V.\,Diwan,}
\author[b]{M.\,J.\,Dolinski,}
\author[c]{A.\,Erickson,}
\author[i]{M.\,Febbraro,}
\author[p]{B.\,T.\,Foust,}
\author[p]{J.\,K.\,Gaison,}
\author[i,k]{A.\,Galindo-Uribarri,}
\author[i,k]{C.\,E.\,Gilbert,}
\author[i,k]{B.\,T.\,Hackett,}
\author[a,1]{S.\,Hans\note{Also at Department of Chemistry and Chemical Technology, Bronx Community College, Bronx, NY, USA.},}
\author[j]{A.\,B.\,Hansell,}
\author[a]{B.\,Hayes,}
\author[p]{K.\,M.\,Heeger,}
\author[b]{J.\,Insler,}
\author[a]{D.\,E.\,Jaffe,}
\author[j]{D.\,C.\,Jones,}
\author[b]{O.\,Kyzylova,}
\author[b]{C.\,E.\,Lane,}
\author[p,2]{T.\,J.\,Langford\note{Also at Yale Center for Research Computing, Yale University, New Haven, CT, 06520.},}
\author[g]{J.\,LaRosa,}
\author[d]{B.\,R.\,Littlejohn,}
\author[i,k]{X.\,Lu,}
\author[d]{D.\,A.\,Martinez\,Caicedo,}
\author[i]{J.\,T.\,Matta,}
\author[o]{R.\,D.\,McKeown,}
\author[e]{M.\,P.\,Mendenhall,}
\author[i]{P.\,E.\,Mueller,}
\author[g]{H.\,P.\,Mumm,}
\author[j]{J.\,Napolitano,}
\author[b]{R.\,Neilson,}
\author[p]{J.\,A.\,Nikkel,}
\author[p]{D.\,Norcini,}
\author[g]{S.\,Nour,}
\author[l]{D.\,A.\,Pushin,}
\author[a]{X.\,Qian,}
\author[i,k]{E.\,Romero-Romero,}
\author[a,3]{R.\,Rosero}
\author[l]{D.\,Sarenac,}
\author[d]{P.\,T.\,Surukuchi,}
\author[g]{M.\,A.\,Tyra,}
\author[i]{R.\,L.\,Varner,}
\author[a]{B.\,Viren,}
\author[d]{C.\,White,}
\author[j]{J.\,Wilhelmi,}
\author[p]{T.\,Wise,}
\author[a]{M.\,Yeh,}
\author[b]{Y.-R.\,Yen,}
\author[a,3]{A.\,Zhang\note{Correspondng authors},}
\author[a]{C.\,Zhang,}
\author[d]{X.\,Zhang}

\affiliation[a]{Brookhaven National Laboratory, Upton, NY, USA}
\affiliation[b]{Department of Physics, Drexel University, Philadelphia, PA, USA}
\affiliation[c]{George W.\,Woodruff School of Mechanical Engineering, Georgia Institute of Technology, Atlanta, GA USA}
\affiliation[d]{Department of Physics, Illinois Institute of Technology, Chicago, IL, USA}
\affiliation[e]{Nuclear and Chemical Sciences Division, Lawrence Livermore National Laboratory, Livermore, CA, USA}
\affiliation[f]{Department of Physics, Le Moyne College, Syracuse, NY, USA}
\affiliation[g]{National Institute of Standards and Technology, Gaithersburg, MD, USA}
\affiliation[h]{High Flux Isotope Reactor, Oak Ridge National Laboratory, Oak Ridge, TN, USA}
\affiliation[i]{Physics Division, Oak Ridge National Laboratory, Oak Ridge, TN, USA}
\affiliation[j]{Department of Physics, Temple University, Philadelphia, PA, USA}
\affiliation[k]{Department of Physics and Astronomy, University of Tennessee, Knoxville, TN, USA}
\affiliation[l]{Institute for Quantum Computing and Department of Physics and Astronomy, University of Waterloo, Waterloo, ON, Canada}
\affiliation[m]{Department of Physics, University of Wisconsin, Madison, Madison, WI, USA}
\affiliation[n]{Physical Sciences Laboratory, University of Wisconsin, Madison, Madison, WI, USA}
\affiliation[o]{Department of Physics, College of William and Mary, Williamsburg, VA, USA}
\affiliation[p]{Wright Laboratory, Department of Physics, Yale University, New Haven, CT, USA}

\emailAdd{azhang@bnl.gov}
\emailAdd{rrosero@bnl.gov} 

\abstract{This work reports the production and characterization of lithium-loaded liquid scintillator (LiLS) for the Precision Reactor Oscillation and Spectrum Experiment (PROSPECT). Fifty-nine 90 liter batches of LiLS (${}^6{\rm Li}$ mass fraction 0.082\%$\pm$0.001\%) were produced and samples from all batches were characterized by measuring their optical absorbance relative to air, light yield relative to a pure liquid scintillator reference, and pulse shape discrimination capability. Fifty-seven batches passed the quality assurance criteria and were used for the PROSPECT experiment.}



\maketitle
\flushbottom
\clearpage 

\section{Introduction}
The Precision Reactor Oscillation and Spectrum Experiment (PROSPECT) is a reactor antineutrino experiment at the Oak Ridge National Laboratory (ORNL)~\cite{ProspectPaper1}. It is deployed to make a precise measurement of the energy spectrum of antineutrinos emitted from the High Flux Isotope Reactor (HFIR) at short baselines in the range  from 7 to 13 meters. 
It also provides a probe of eV-scale sterile neutrino oscillations.

The PROSPECT detector is a $\sim\!2.0$ m \texttimes~1.6 m \texttimes~1.2 m active rectangular volume containing about four tons of non-flammable liquid scintillator loaded with $^{6}$Li (LiLS) to a 
mass fraction of approximately $0.1\%$~\cite{PROSPECT_NIMpaper}. 
The detector operates with minimal overburden from the HFIR building. Thin specularly reflecting panels divide the LiLS volume into an 11\texttimes14 array of 154 optically isolated rectangular segments (14.5 cm \texttimes~14.5 cm \texttimes~117.6 cm each) viewed on both ends by 5-inch (12.7 cm) photomultiplier tubes (PMTs).

Including a 20\% contingency, the production of approximately 5000 liters LiLS was required. LiLS was chosen because 
it allows efficient capturing of neutrons produced by the inverse beta decay (IBD) reaction in a compact detector.
$^{6}$Li has a high capture cross section for IBD neutrons, and an alpha and a triton are produced from the capture with approximately 540 keV of visible energy in the scintillator~\cite{Li6Characterization, LiLS_paper}. 
These particles are spatially localized, and the deposited energy can be distinguished from electron-like events by 
pulse shape discrimination (PSD).

The collaboration developed a LiLS that fulfills the requirements of the experiment in terms of light yield, PSD, and stability. The chosen LiLS formulation is the result of about three years of development. The production of LiLS was carried out at the Chemistry Department of Brookhaven National Laboratory (BNL). Fifty-nine 90 liter batches of LiLS were produced over a period of nine months, starting in Jan. 2017. 

This paper summarizes the LiLS production and its quality assurance (QA) program. Criteria for acceptance were based on comparison to the performance of an initial LiLS batch deployed in a 50 liter, two-segment prototype detector known as PROSPECT-50~\cite{P50paper}. PROSPECT-50 was designed to replicate essential properties of the 154-segment PROSPECT and achieved PSD and light yield of the LiLS satisfying the PROSPECT design requirements. Each LiLS production batch was required to be consistent with or superior to that of PROSPECT-50 in terms of absorbance at 420 nm of incident light relative to air, light yield relative to a liquid scintillator reference (linear alkylbenzene, LAB), and PSD.

\section{LiLS production}

\subsection{Materials and quality assurance}

The generic formula of the LiLS consists of a nonionic surfactant, a 9.98 mol/L aqueous solution of lithium chloride (LiCl) with 95\% enriched ${}^{6}{\rm Li}$ by atom, 2,5-diphenyloxazole (PPO) and 1,4-bis(2-methylstyryl) benzene (bis-MSB) in a commercial, di-isopropylnaphthalene (DIN)-based scintillator (EJ-309)\footnote{Certain trade names and company products are mentioned in the text or identified in illustrations in order to adequately specify the experimental procedure and equipment used. In no case does such identification imply recommendation or endorsement by the National Institute of Standards and Technology, nor does it imply that the products are necessarily the best available for the purpose.}. The surfactant is an ether-based glycol from DOW Chemical. The PPO and bis-MSB were obtained from Research Product International.

\begin{figure}[htb]
  \centering
  \includegraphics[width=12cm]{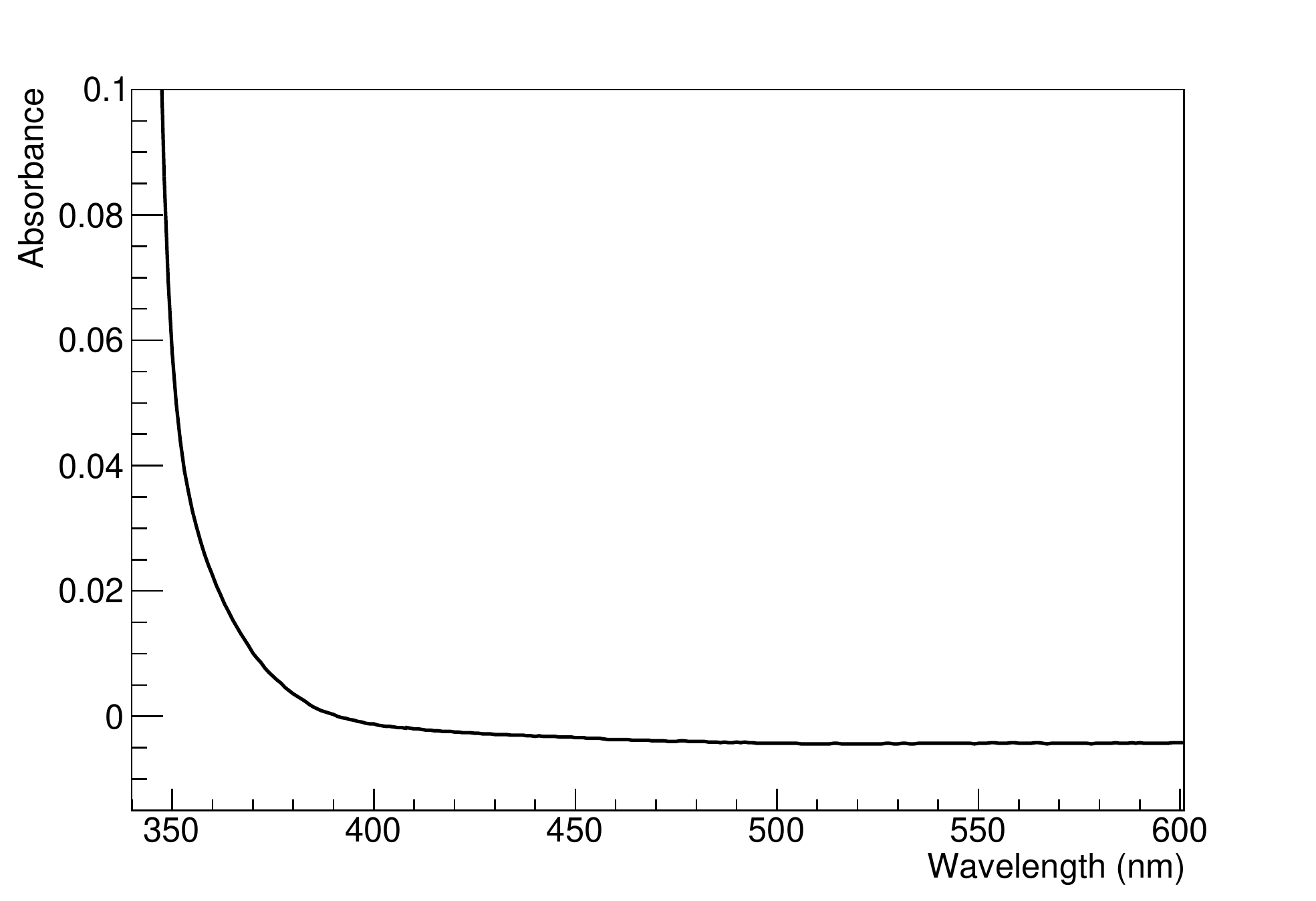}
  \vspace{-0.5cm}
  \caption{The relative absorbance of a raw EJ-309 scintillator sampled from drum 10.}
  \label{absEJ309}
\end{figure}

\begin{figure}[htb]
  \centering
  \includegraphics[height=7.5cm]{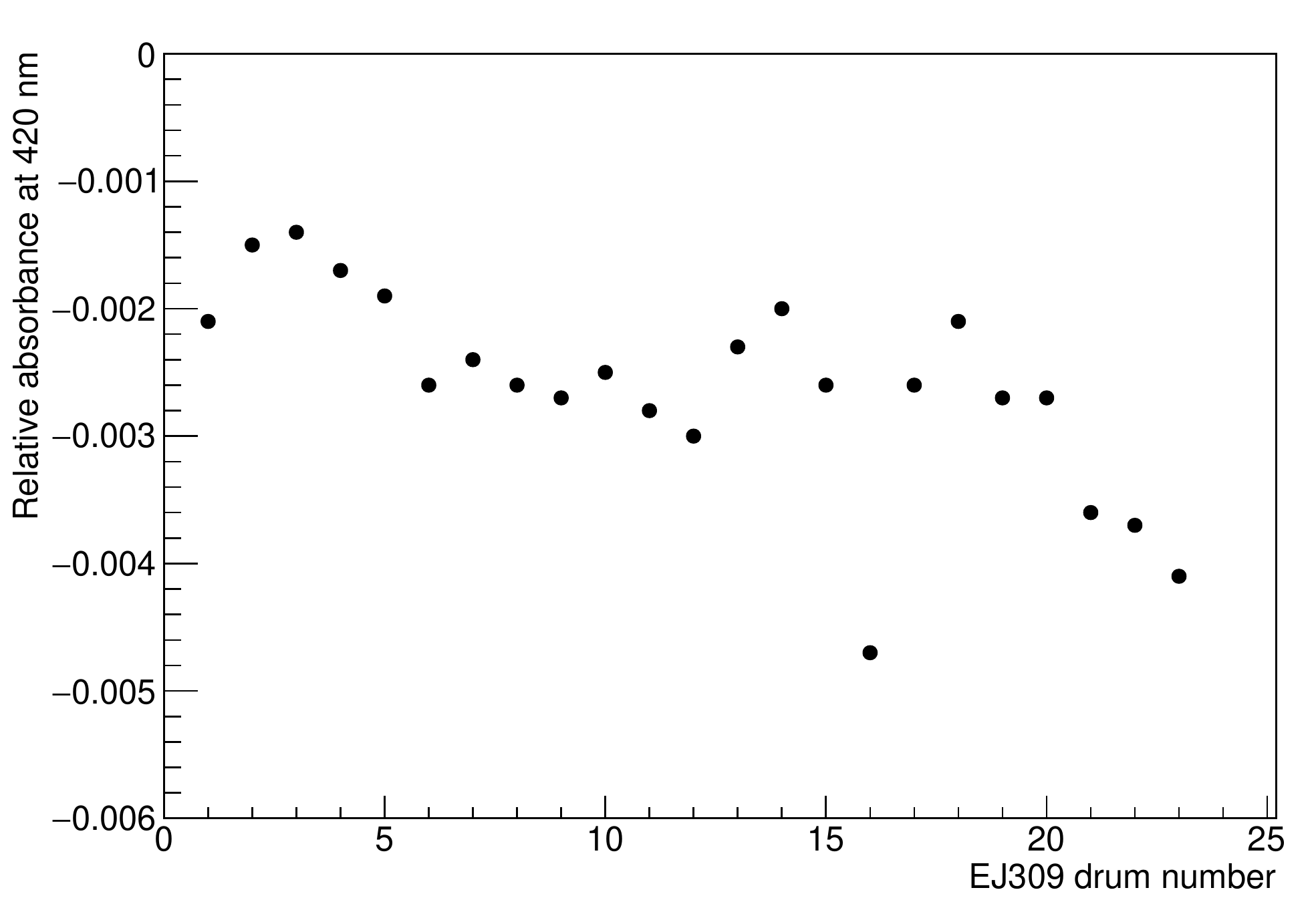}
  \vspace{-0.5cm}
  \caption{The relative absorbance at 420 nm of EJ-309 scintillator from different drums.} 
  \label{absEJ309_420nm}
\end{figure}

The EJ-309 scintillator was purchased from Eljen Technology, it was delivered in 23 drums (4600 L in total). The absorbance of a sample of each EJ-309 drum relative to air was measured using a UV-vis spectrophotometer (Shimadzu UV-1800). Two cylindrical quartz glass cells (10 cm path length, 25.4 mm diameter, 1 mm wall thickness) were used in the measurements, with one holding a 48 mL liquid sample and the other left empty for reference. Each sample's absorbance spectrum was measured by scanning from wavelength 200 nm to 1100 nm in 1 nm steps. Figure~\ref{absEJ309} shows the measured relative absorbance spectrum of an EJ-309 scintillator sample from drum 10 over the wavelength range from 350 nm to 600 nm where the PMTs are sensitive. Negative values in absorbance are attributed to the refractive index differences between the sample (scintillator) and reference (air) in the measurements. Figure~\ref{absEJ309_420nm} shows relative absorbance of all EJ-309 samples at 420 nm, near the most sensitive wavelength for the PMTs. Drums 2-4 had a slightly higher absorbance due to oxygen contamination in the manufacturer's nitrogen purging line. 

The LiCl solution was supplied by the National Institute of Standards and Technology from enriched lithium carbonate (95$\pm$1\% $^{6}$Li by atom) and analytical grade concentrated (37$\pm$1\% by mass) hydrochloric acid, produced with the chemical reaction 

\begin{equation}
 {\rm Li}_{2}{\rm CO}_{3}+2{\rm HCl}\rightarrow2{\rm LiCl}+{\rm H}_{2}{\rm O}+{\rm CO}_{2}.
\end{equation}

\begin{figure}[htb]
  \centering
  \includegraphics[width=12cm]{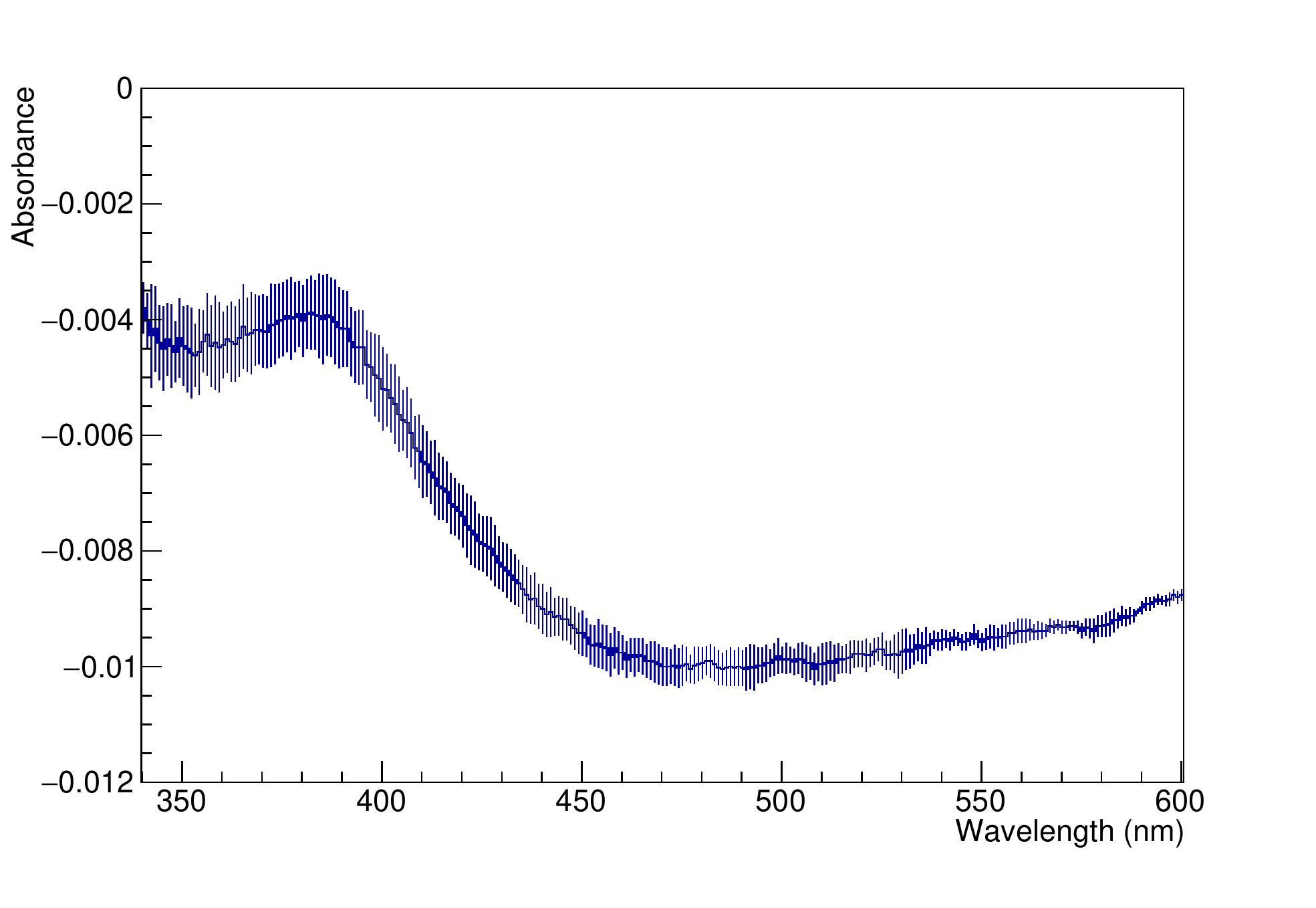}
  \vspace{-0.5cm}
  \caption{The average absorbance curve for the LiCl solution. 
  The error bars represent the standard deviation calculated from the measured samples at each wavelength.}
  \label{absLiCl}
\end{figure}

The LiCl solution was filtered and passed through an anion exchange chromatography column (Bio-Rad AG 1-X4, 100 to 200 mesh) to remove colored (e.g., iron) impurities. A total of 86 L of 9.98$\pm$0.02 mol/L purified LiCl solution was prepared in six individual batches. The average absorbance of the purified LiCl batches is shown in Figure~\ref{absLiCl}.

\subsection{LiLS production procedures and storage}

\begin{figure}[htb]
  \begin{center}
       \vspace{-0.1cm}
          \begin{minipage}[b]{0.3\textwidth}
            \centering
            \includegraphics[width=4.5cm, height=7.5cm]{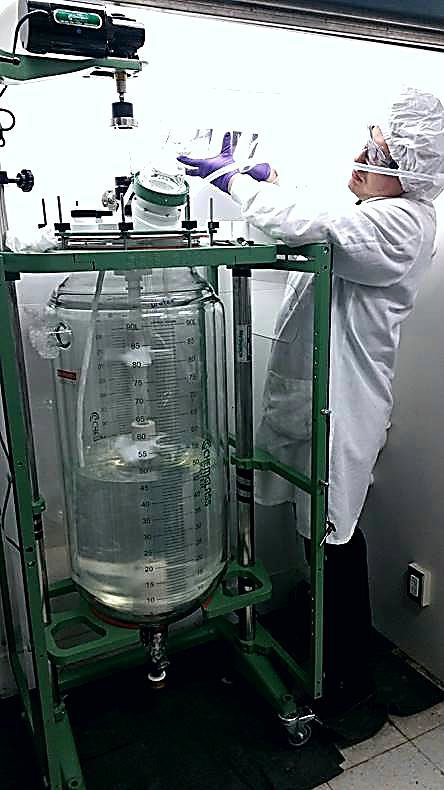}
          \end{minipage}%
          \begin{minipage}[b]{0.3\textwidth}
            \centering
            \includegraphics[width=4.5cm, height=7.5cm]{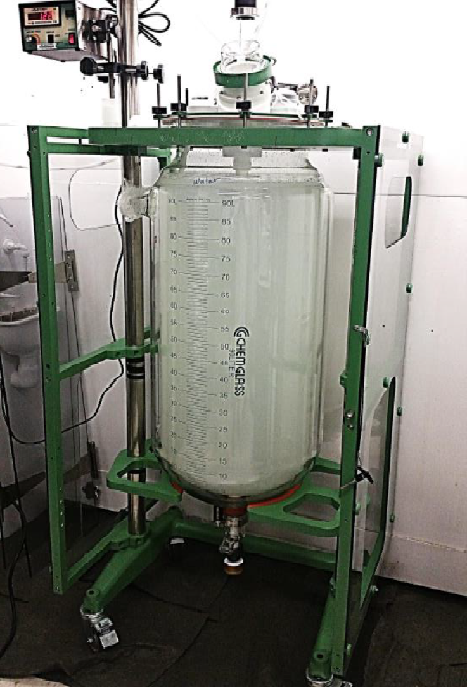}
          \end{minipage}%
          \begin{minipage}[b]{0.3\textwidth}
            \centering
            \includegraphics[width=4.5cm, height=7.5cm]{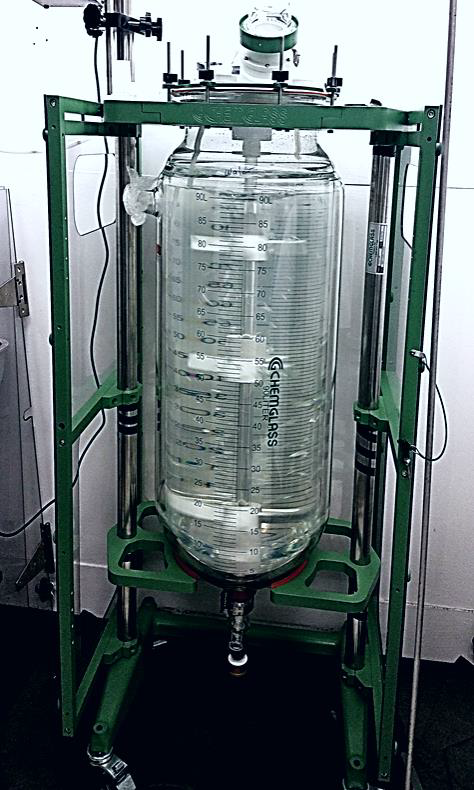}
          \end{minipage}%
        \vspace{-0.3cm}
        \caption{LiLS production system at different stages in the 90 L Chemglass: injecting of raw material into the reactor ($\it{left}$), mixing in progress ($\it{middle}$) and synthesis completion ($\it{right}$). The liquid is opaque during the synthesis and becomes clear after synthesis is done.}
        \label{90Lreactor}
  \end{center}
\end{figure}

The LiLS was produced by first purifying the raw components and then mixing in stages in a double-jacketed 90 L Chemglass reaction vessel, shown in Figure~\ref{90Lreactor}. The metal-loading principle of water-based liquid scintillator (WbLS) was applied to the synthesis of LiLS~\cite{WbLStechnique}. 
Since WbLS loading is a one-step, direct aqueous/organic mixing procedure, as shown in Figure~\ref{90Lreactor}, no loss of $^{6}$Li during synthesis in LS is expected.
The fraction of $^{6}$Li loaded in the scintillator was  $(0.082 \pm 0.001)\%$ by mass~\cite{Li6fraction}. 

The ether-based surfactant was purified by thin-film vacuum distillation while the EJ-309 was pre-purified by the manufacturer. The reaction vessel has several injection ports made of polytetrafluoroethylene (PTFE) for adding chemical materials. All the tubing, filtration system, liners, and the mixing system were pre-cleaned with ethanol (Ethyl alcohol 190 proof), rinsed with ultrapure water (resistivity 18.2 M\textohm$\times$cm) and dried with nitrogen gas. The system was sealed in an inert environment until use. 

All purified raw materials were introduced through different ports into the reactor at different synthesis stages and mixed for 2 hours. After the synthesis was completed, the LiLS was discharged through a 2 $\mu$m glass filter (Whatman) situated in a 316-stainless steel filtration system and stored in a 208 L lined drum. Each drum was equipped with a 5 $\mu$m thick perfluoroalkoxy alkane inner bag for liquid storage and a 5 $\mu$m thick outer polypropylene liner for secondary containment. 
The maximum storage capacity of each drum was limited to 180 L, equal to two 90 L batches, for ease of handling and overflow prevention. 
Nitrogen cover gas was added to the rest of the volume. 
All materials  in contact with the LiLS are compatible with LiLS. 
During batch production, two one-liter-sized samples from every batch were taken for quality assurance measurements.

During the production period, one to three batches were produced weekly with 57 batches produced in the first six months. The production of the final two batches was delayed by three months due to a temporary shortage of raw scintillator. Batches 1 and 2 were used for prototyping and material compatibility tests. All other drums were stored in a temperature-controlled warehouse at BNL and later transported in temperature-controlled trucks to ORNL for the PROSPECT experiment. The temperature was controlled between 20 \textcelsius \,and 30 \textcelsius.

\section{LiLS quality assurance measurements}

In this section, the measurement methods, data analysis strategies, selection criteria, and results are described for qualifying the LiLS samples in terms of their optical absorbance, light yield, and PSD. 

\subsection{Relative optical absorbance}

\begin{figure}[h]
  \centering
  \includegraphics[width=12cm]{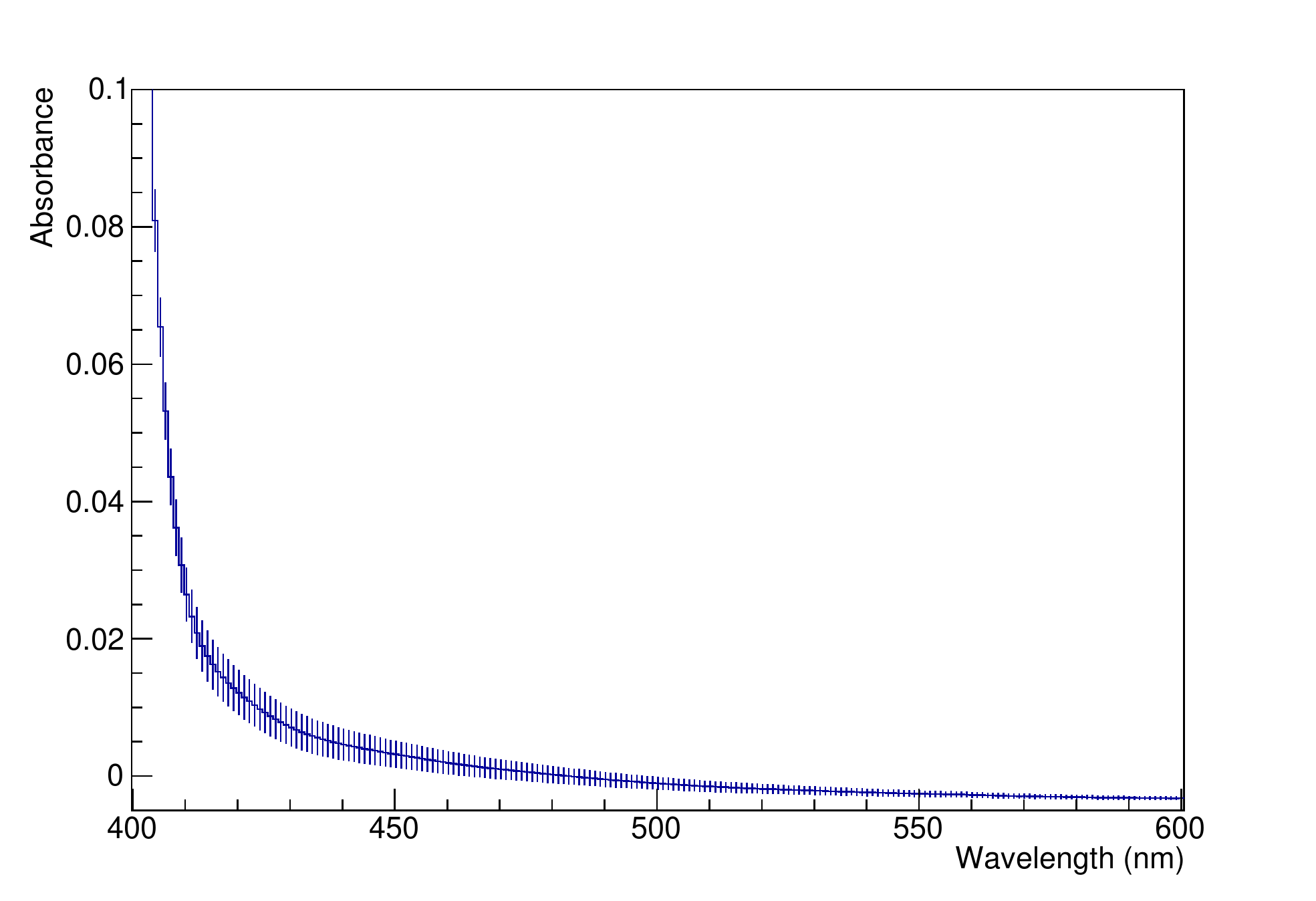}
  \vspace{-0.3cm}
  \caption{Average of the relative absorbance of all the measured LiLS samples. The error bars represent the standard deviation calculated from the measured samples at each wavelength.}
  \label{absLiLS}
\end{figure}

\begin{figure}[h]
  \centering
  \includegraphics[height=8cm]{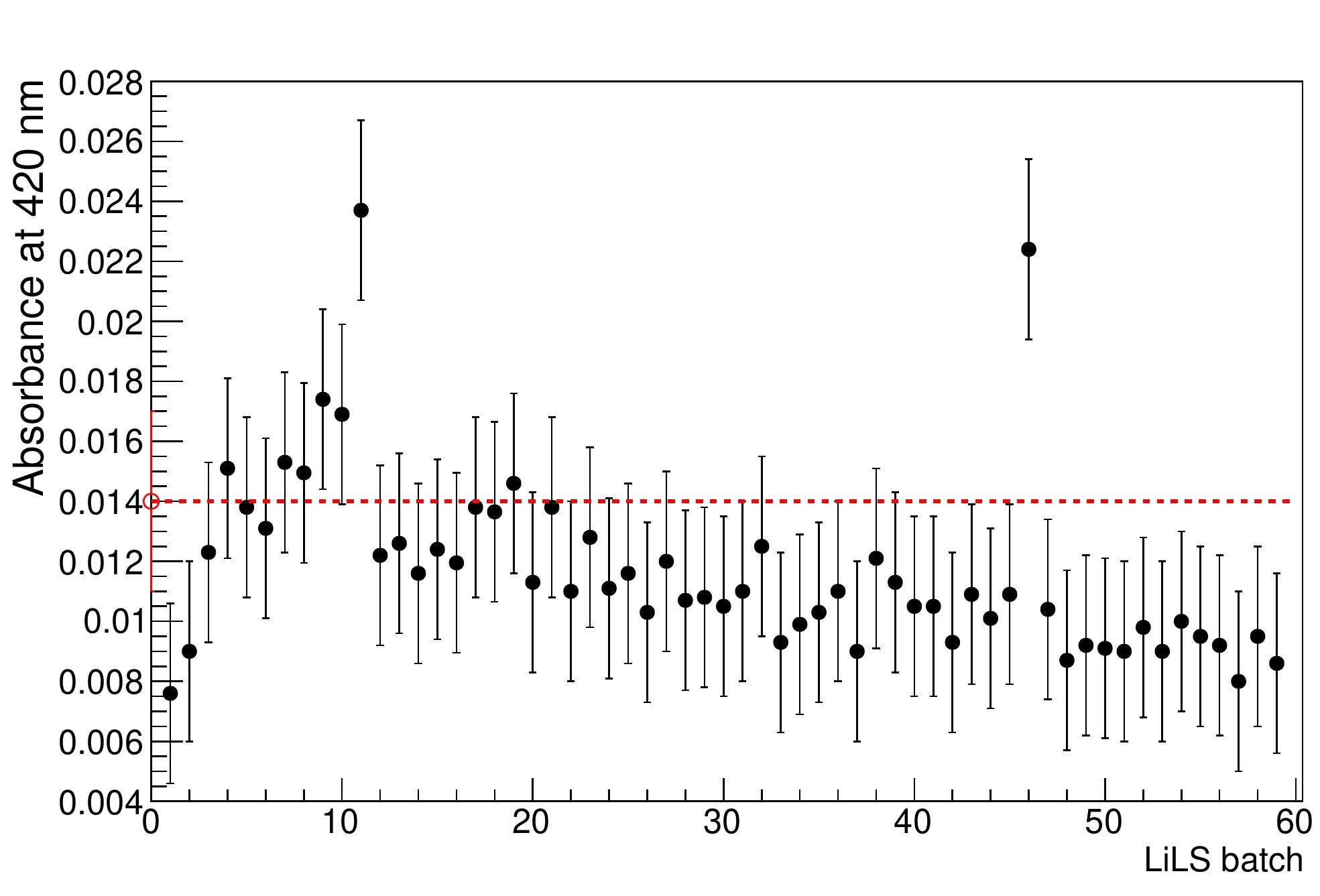}
  \vspace{-0.5cm}
  \caption{Relative absorbance at 420 nm for every measured LiLS sample (black points) compared to the PROSPECT-50 sample indicated by the red dashed line. Batches 11 and 46 were rejected due to their high absorbance. A common 0.003 uncertainty is assigned to each point as described in the text.}
  \label{absLiLS_420nm}
\end{figure}

The relative optical absorbance of a sample from each batch was measured immediately after production. Figure~\ref{absLiLS} shows average absorbance for all the measured samples. Figure~\ref{absLiLS_420nm} presents the samples' absorbance at 420 nm, the absorbance for a PROSPECT-50 sample is included for comparison. A standard deviation of 0.003 for UV absorbance was determined from measurements of nine PROSPECT-50 samples and assigned as the systematic uncertainty of each sample measurement. 
The large variation in the first 10 batches is attributed to the oxygen contamination of EJ-309 as described earlier. 
The absorbance for batches 11 and 46 was clearly higher than other samples and therefore they were rejected. The other batches were all considered acceptable with absorbance comparable to the PROSPECT-50 sample. The rejected samples were not subjected to light yield and PSD measurements.

The optical absorbance of the LiLS was stable from production to deployment. 
The absorbance of repeated measurements of samples from a stored drum taken over a six month period showed no variations greater than the estimated 0.003 systematic uncertainty 
in the 400 nm to 600 nm range. 
Samples from drums after shipment from BNL to ORNL show similar behavior. 
All drums of LiLS shipped to ORNL were accepted for deployment in the PROSPECT detector.

\begin{figure}[htb]
  \centering
  \includegraphics[height=8cm]{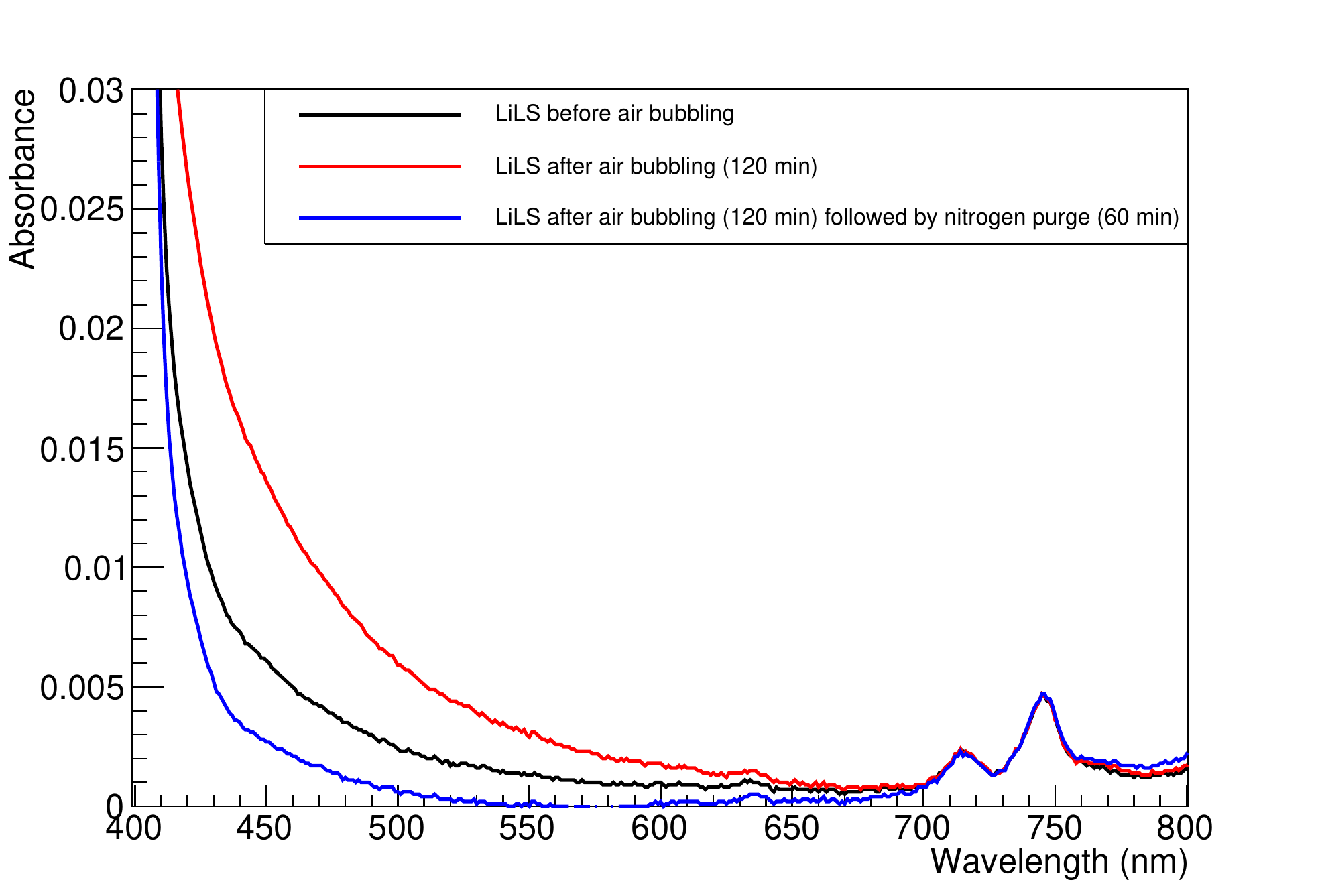}
  \vspace{-0.5cm}
  \caption{The relative absorbance spectra of LiLS samples before and after exposure to air, and subsequent nitrogen sparging demonstrating the oxygen quenching effect in the LiLS. The gas flow rates for air and nitrogen were set at 30 mL/min.}
  \label{LiLS_Quenching}
\end{figure}

A measurement of the oxygen quenching effect on absorbance is shown in Figure~\ref{LiLS_Quenching}. The LiLS absorbance degrades when oxygen is introduced by bubbling air through the scintillator. Sparging the LiLS with nitrogen gas removes the oxygen and improves the absorbance~\cite{LABquenching}. To minimize quenching, all the LiLS samples were sparged with nitrogen gas at 30 mL/min for 30 min prior to the light yield measurements. 

\begin{figure}[htb]
  \centering
  \includegraphics[width=12cm]{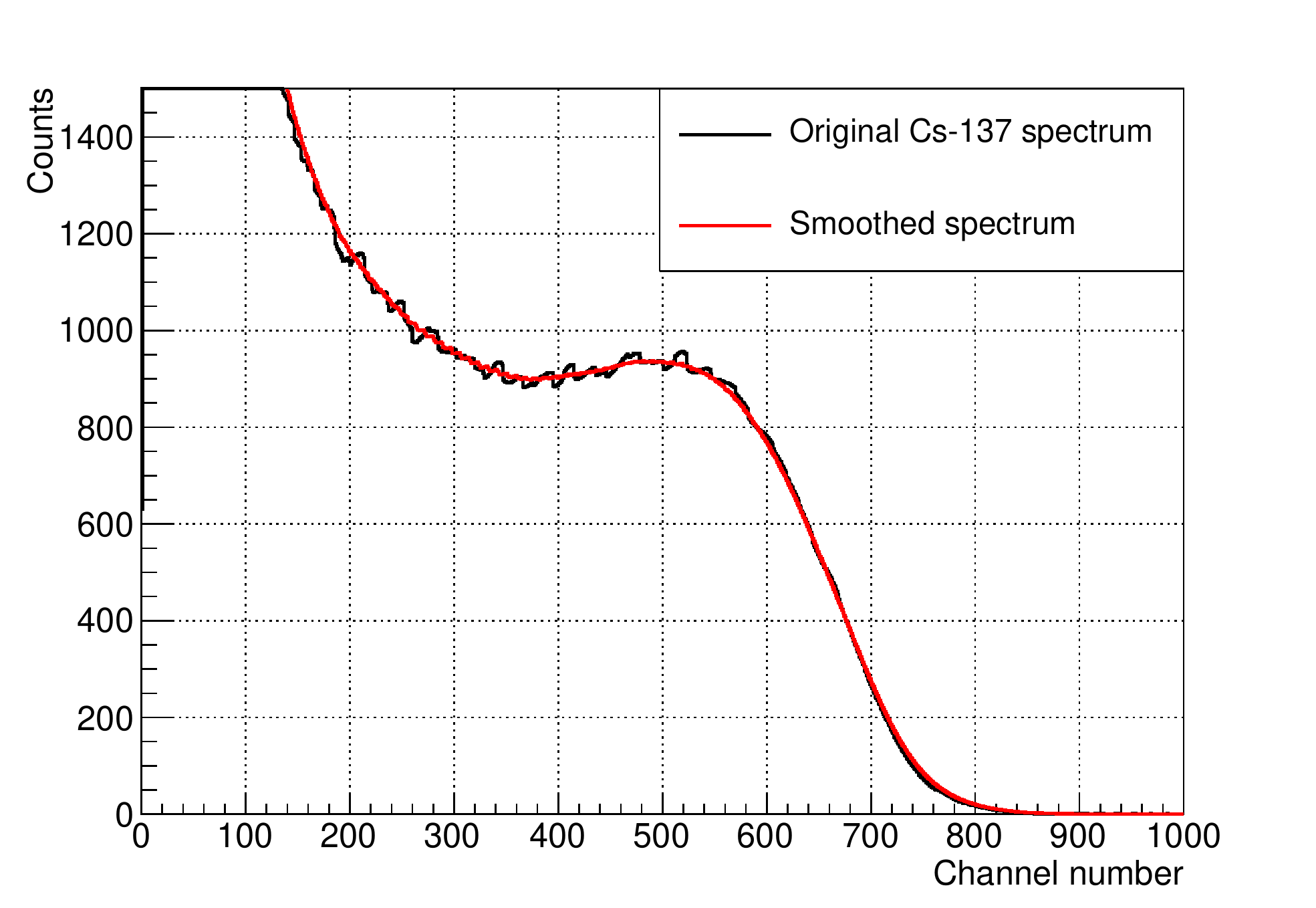}
  \vspace{-0.3cm}
  \caption{The measured $^{137}$Cs spectrum by the LS6500 for the LAB reference with the original spectrum (in black) and the smoothed distribution used for determining the Compton edge (in red).}
  \label{LABspectrumEg}
\end{figure}

\subsection{Relative light yield}

The light yield of a LiLS sample from each 90 L batch relative to a LAB sample was measured with an automatic liquid scintillator counter (Beckman LS6500). The counter is equipped with a $^{137}$Cs gamma source, two photomultiplier tubes in coincidence mode, and a multichannel analyzer.

Liquid scintillator samples of about 10.6 g each were filled in borosilicate glass vials with white polyethylene caps with PTFE inner liners (vial outer diameter 28.6 mm and height 61 mm). The LiLS samples were sparged with nitrogen at $\sim$30 mL/min for about 30 minutes and then closed for measurement. No effort was made to remove air from the closed vials, and the light yield subsequently diminished due to oxygen quenching~\cite{LABquenching, QuenchingRef}.

The counter enabled automatic cyclical measurements of the reference and LiLS samples over an extended period of typically 70 hours. 
Each sample was measured for 15 minutes in each cycle. 
Three quantities were extracted from the data for each LiLS sample for the QA: the initial light yield relative to the LAB reference and the fast and slow time constants. 
The initial relative light yield is the main quantity used for LiLS acceptance.
The fast time constant is attributed to oxygen quenching and the slow time constant indicates the LiLS performance stability. 

There are artifacts in the observed spectrum that arise from the LS6500 counter~\cite{LBignell}, as can be seen in Figure~\ref{LABspectrumEg}. 
The original spectrum is smoothed by applying the Gauss-Hermite quadrature and the smoothed spectrum is used to extract the Compton edge as a measure of the light yield following the differentiation method described in~\cite{differentiationmethod}.

\begin{figure}[htb]
  \begin{center}
       \vspace{-0.3cm}
        \hspace{-1cm}
          \begin{minipage}[b]{0.5\textwidth}
            \centering
            \includegraphics[width=9cm, height=5cm]{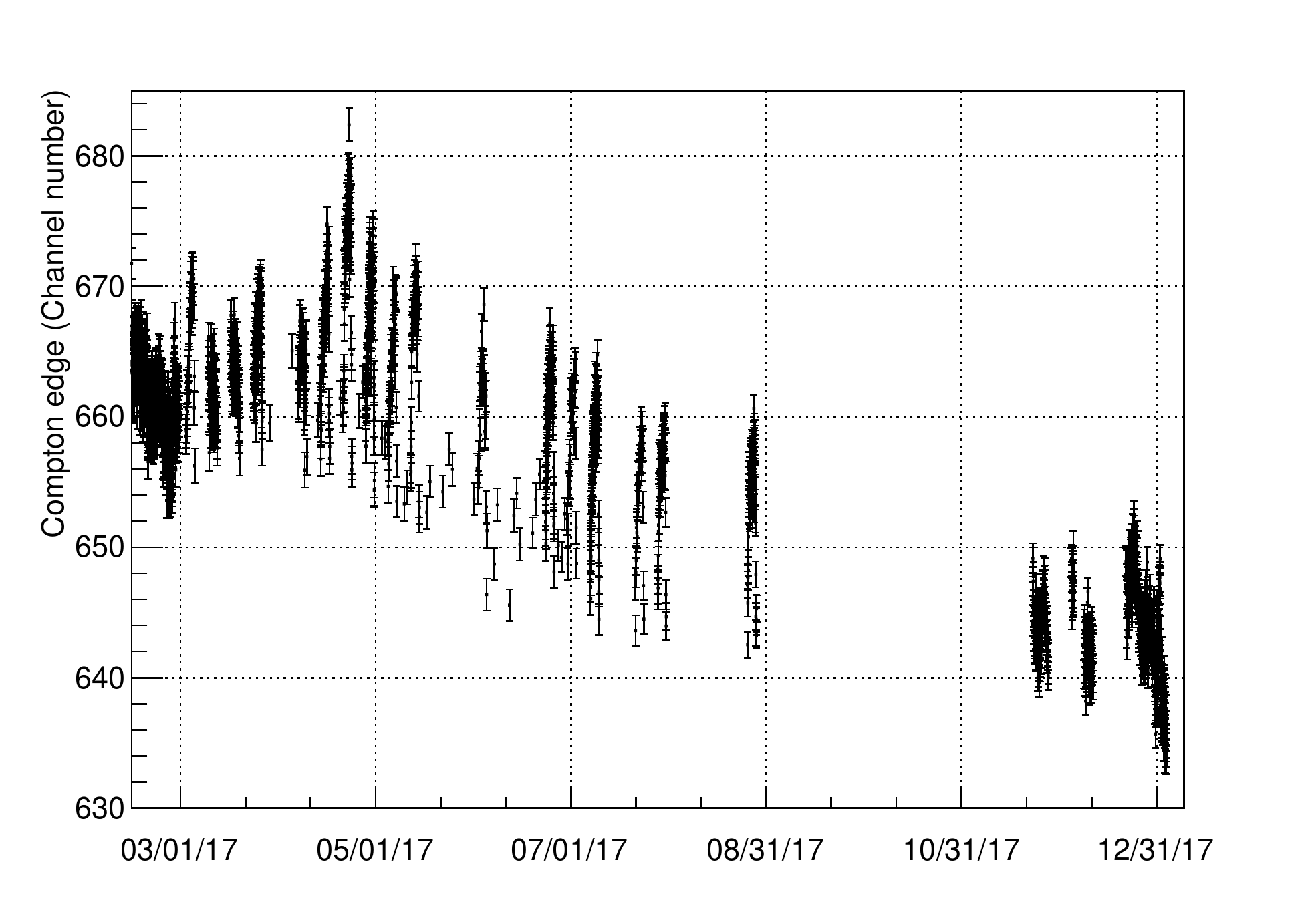}
          \end{minipage}%
          \begin{minipage}[b]{0.5\textwidth}
            \centering
            \includegraphics[width=6cm, height=5cm]{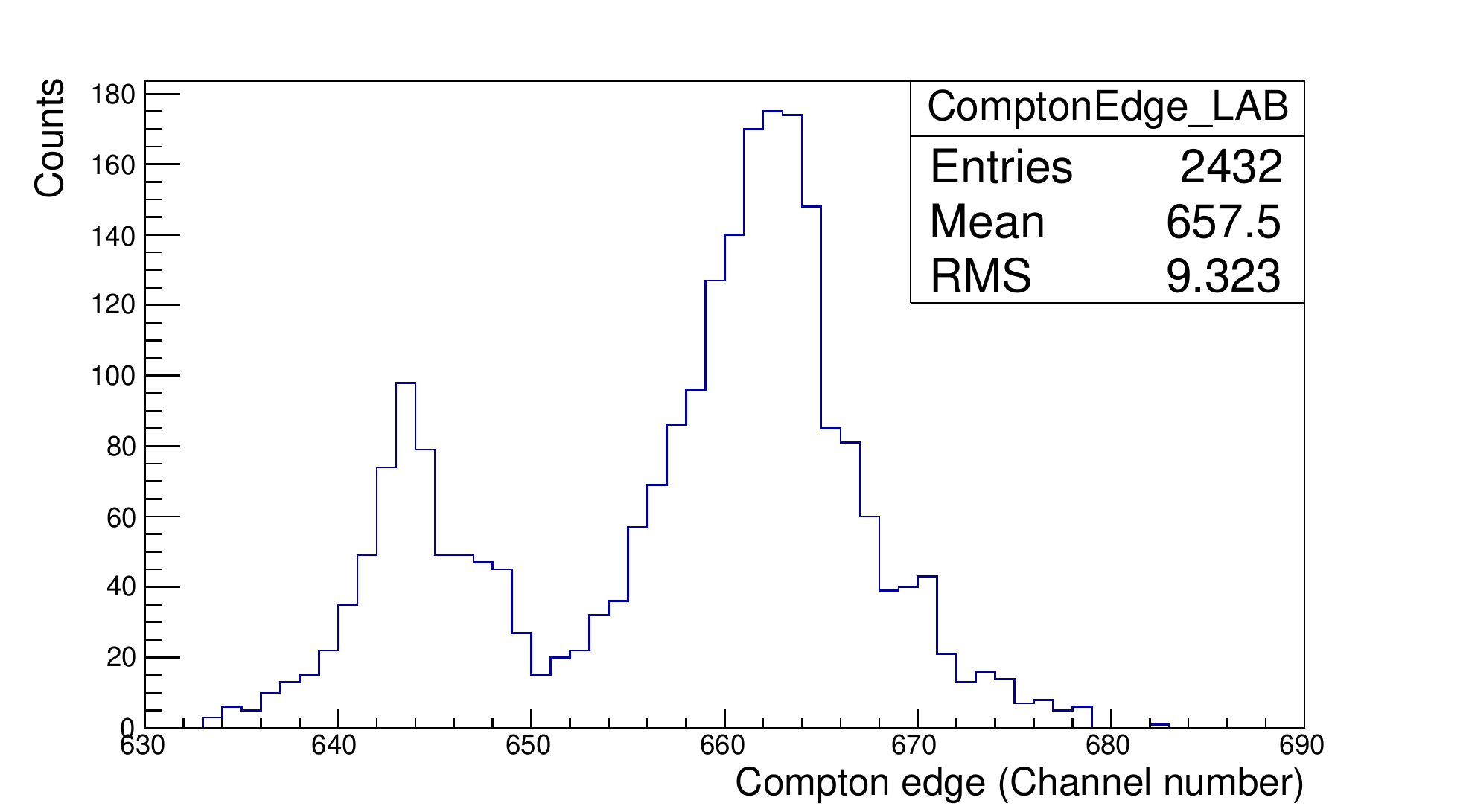}
          \end{minipage}%
        \vspace{-0.3cm}
        \caption{The measured LAB Compton edge over a period of about 10 months ($\it{left}$) and its distribution ($\it{right}$). 
        The error bars on the left are statistical.}
        \label{LAB_ComptonEdge}
  \end{center}
\end{figure}

The identified Compton edge of the reference LAB sample used for the quality assurance measurements as a function of time is shown in Figure~\ref{LAB_ComptonEdge}. 
The variation in the reference sample light yield is consistent with the long term behavior of the LS6500 observed previously~\cite{LBignell}. 

\begin{figure}[htb]
  \begin{center}
       \vspace{-0.3cm}
       \hspace{-1cm} %
          \begin{minipage}[b]{0.5\textwidth}
                      \centering
            \includegraphics[width=7.5cm, height=6cm]{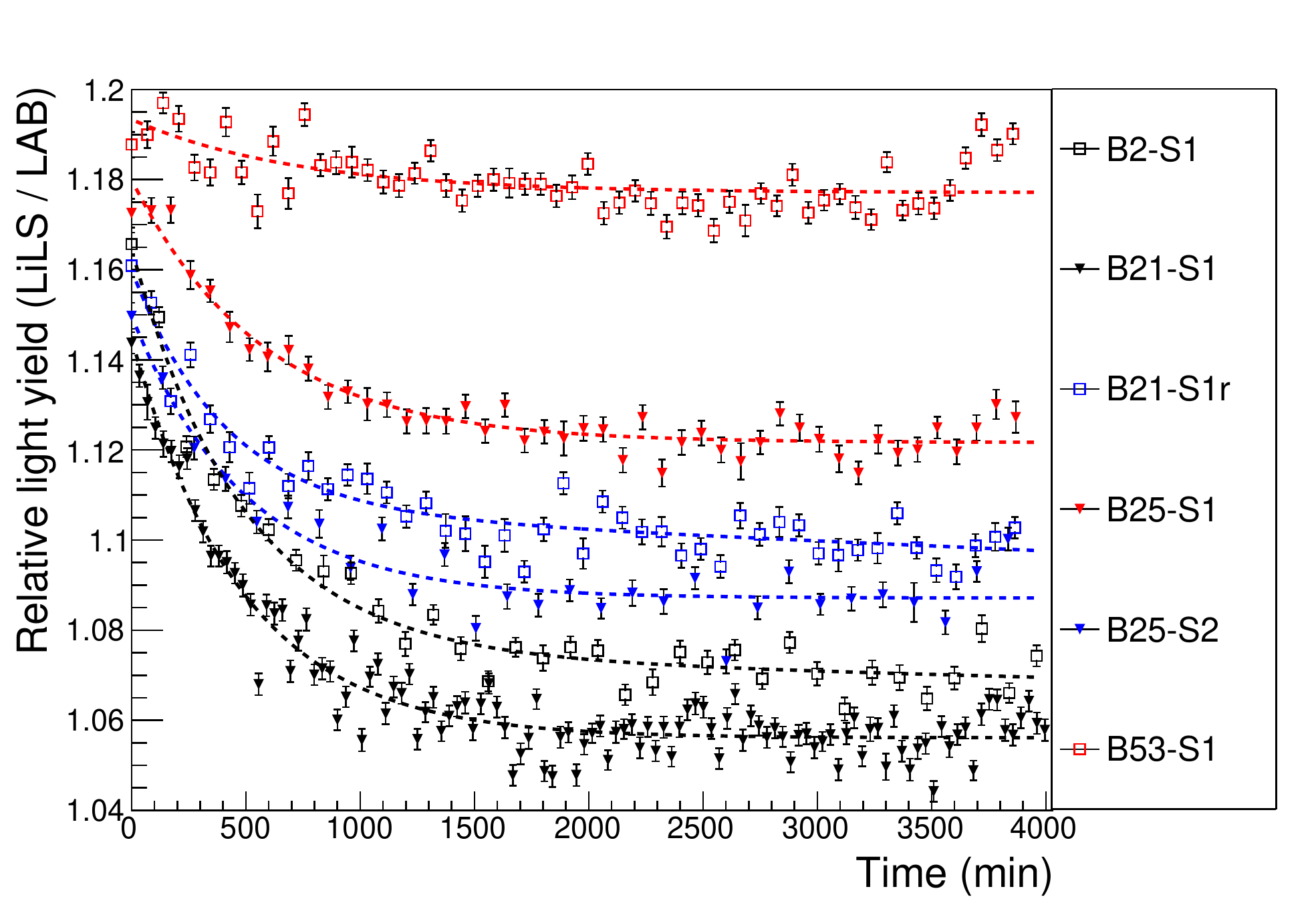}
          \end{minipage}
          \begin{minipage}[b]{0.5\textwidth}
            \centering
            \includegraphics[width=8cm, height=6cm]{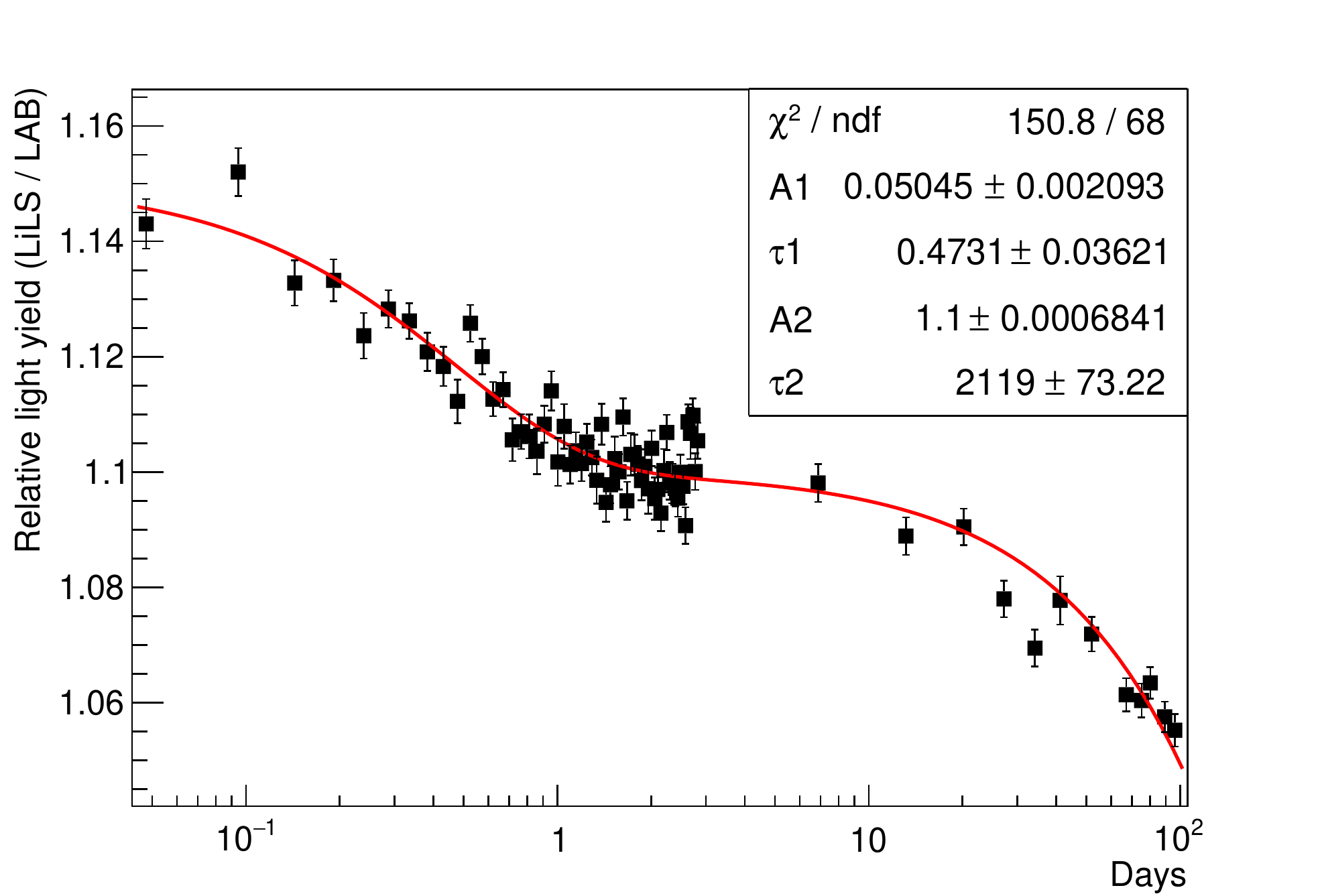}
          \end{minipage}%
        \vspace{-0.3cm}
        \caption{$\it{Left}$: Relative light yield of a few representative LiLS samples during the $\sim$70 hours of measurements. ``B'' and ``S'' denote batch and sample number respectively. Occasionally multiple samples from the same batch were measured (B25-S1 and B25-S2) and the same sample was re-measured (B21-S1r). Each sample's light yield is fitted with two exponentials (dashed lines) as described in the text. $\it{Right}$: The relative light yield of the sample B18-S1 over 100 days. The fitted parameters are shown on the plot.
        }
        \label{LiLS_LY_samples}
  \end{center}
\end{figure}

Each LiLS sample's light yield was assessed using $\sim$70 hours data. 
The relative light yield is defined by the ratio of Compton edges between the LiLS sample and LAB reference.
Figure~\ref{LiLS_LY_samples} shows the relative light yield of representative LiLS samples in the first 70 hours. 
All other samples have a light yield between the extreme curves shown in Figure~\ref{LiLS_LY_samples}. 
The relative light yield decreases rapidly in the first $\sim$500 minutes and then undergoes a much slower decrease. 
Consequently, we fit this time-dependence with a double exponential function 
\begin{equation}
f(t) =
A_{1} e^{-t/\tau_{1}}+A_{2} e^{-t/\tau_{2}}\ \ , 
\end{equation}
\noindent in which $A_{1,2}$ and $\tau_{1,2}$ are the corresponding relative light yields and time constants for the fast and slow components, respectively. 
The sum $A_{1}+A_{2}$ describes the initial relative light yield of an LiLS sample and is the main quantity that is used in the QA process. 
Some samples were monitored for a few months and the fitted $\tau_{2}$ is at least four years. 
As an example, the fitted measurements 
for the 
batch 18 sample are shown in Figure~\ref{LiLS_LY_samples}.
\begin{figure}[h]
  \begin{center}

       \vspace{-0.3cm}
       \hspace{-6cm} %
          \begin{minipage}[b]{0.5\textwidth}
                      \centering
            \includegraphics[width=16cm]{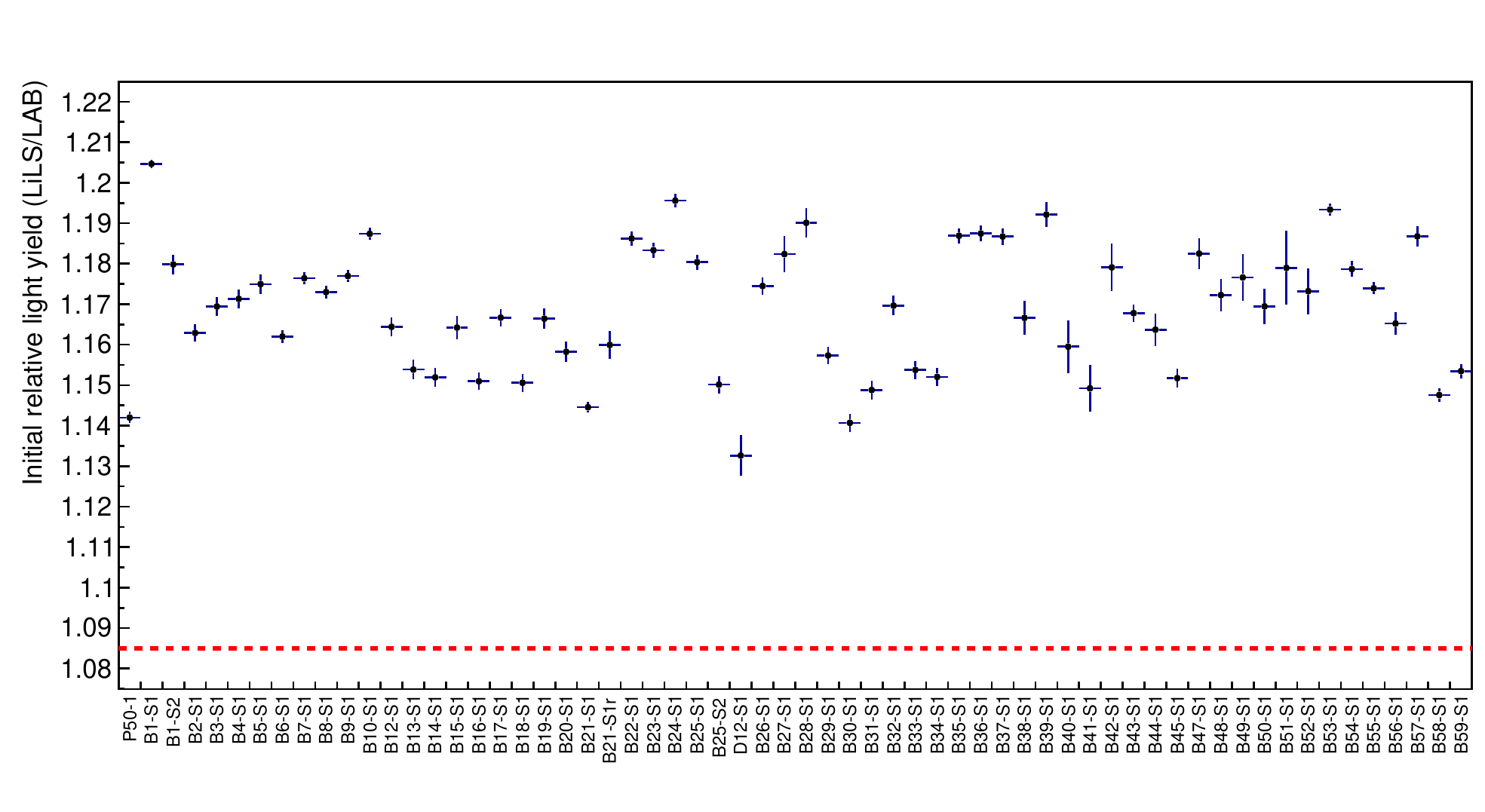}
          \end{minipage}\newline
          \begin{minipage}[b]{0.5\textwidth}
            \centering
            \includegraphics[width=8cm]{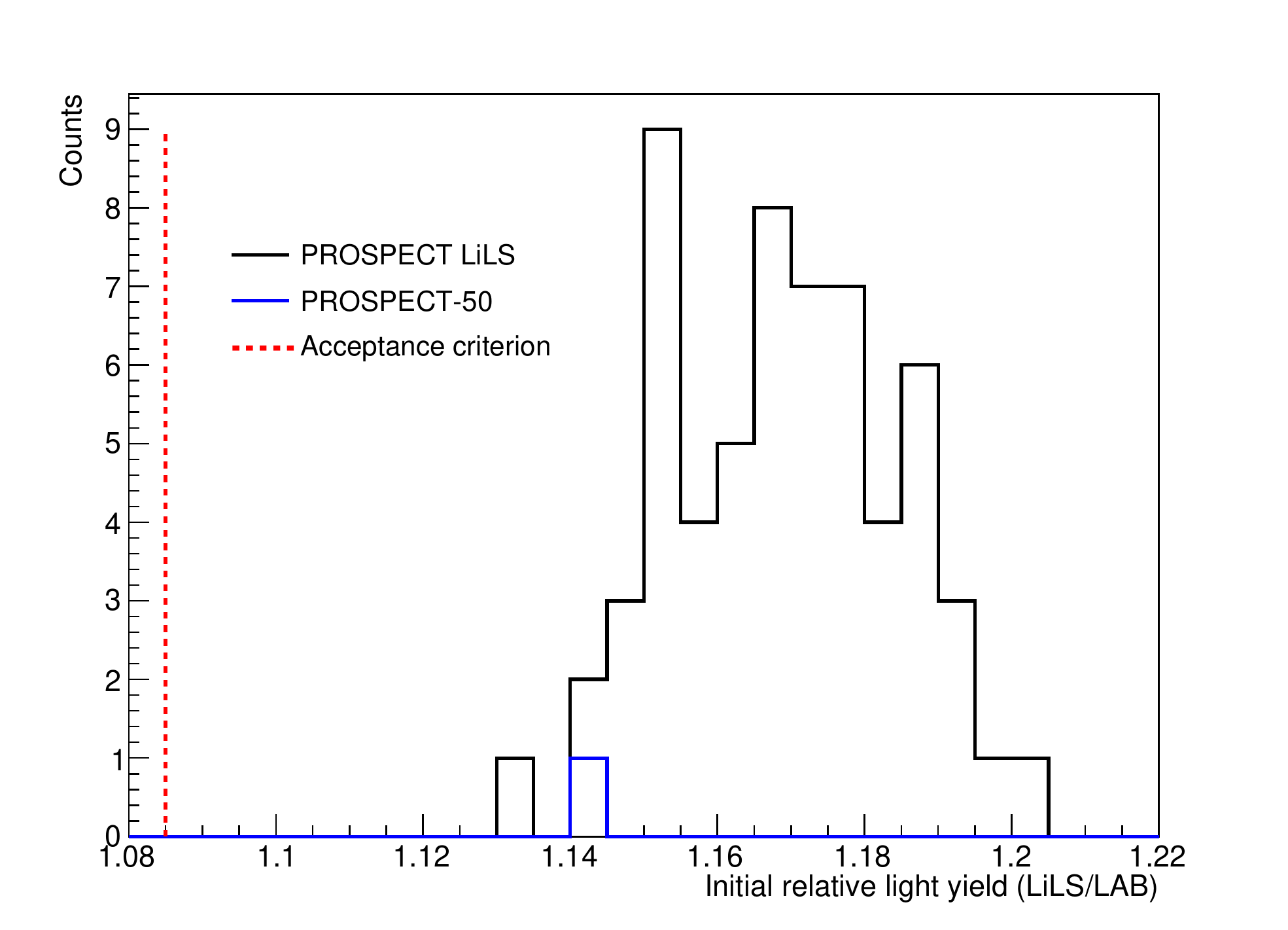}
          \end{minipage}%

        \caption{
        Initial relative light yield  of each 
        LiLS sample ($\it{top}$) and its distribution 
        ($\it{bottom}$) 
        from the Compton edge from a $^{137}$Cs source 
        as described in the text.
        Batch 25 was measured twice (B25-S1 and B25-S2). 
        The sample for batch 21 was remeasured (B21-S1r).
        }
        \label{LiLS_LY_samples2}
  \end{center}
\end{figure}

Figure~\ref{LiLS_LY_samples2} shows the initial relative light yield for all the LiLS samples as well as a sample from PROSPECT-50~\cite{P50paper}. 
The PROSPECT-50 sample was used to set the light yield selection criterion: the initial relative light yield of LiLS was required to be greater than 95\% of the PROSPECT-50 sample; that is, $A_{1}+A_{2}>1.085$ since the measured initial light yield for the PROSPECT-50 sample was 1.142.
As can be seen from Figure~\ref{LiLS_LY_samples2}, all the measured samples show a satisfactory initial light yield well above the required threshold. 


\subsection{Pulse shape discrimination}

The same LiLS samples were used in the PSD measurements right after the light yield measurement without additional nitrogen sparging.
The PSD capability of the LiLS samples was measured using an $^{241}$Am-$^{9}$Be neutron source (105 MBq) from which the rate of nuclear recoils in a $\sim$10 g sample largely exceeded the thermal neutron capture rate. The PSD for nuclear recoils as a function of visible energy is measured and the result is used to estimate the PSD for thermal neutron captures according to the measured response of the LiLS in reference~\cite{P50paper}. 

\begin{figure}[htb]
  \centering
  \includegraphics[width=7cm]{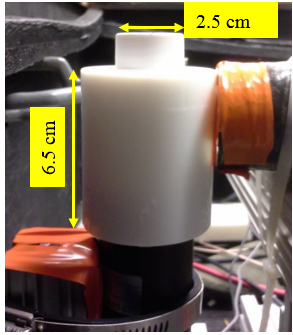}
  \caption{Side view of the setup for the PSD measurements. The PMT is vertically mounted and facing up, the vial with LiLS rests on top of the PMT and it is surrounded by a PTFE cylinder to enhance light collection. The Am-Be neutron source shown mounted at the right side of the PTFE cylinder.}
  \label{PSDsetup}
\end{figure}

\begin{figure}[htb]
  \centering
  \includegraphics[width=12cm]{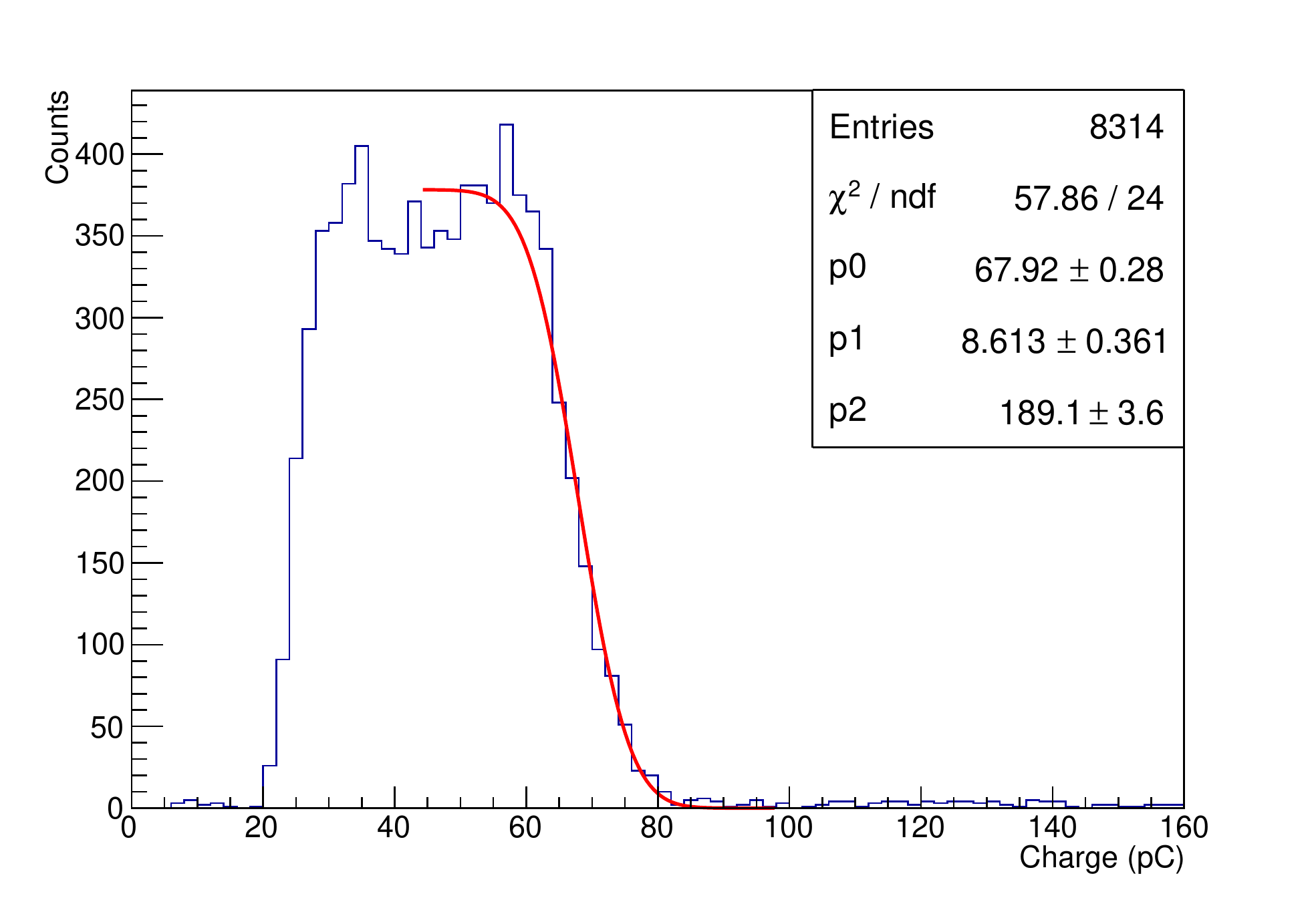}
  \vspace{-0.3cm}
  \caption{$^{137}$Cs spectrum of the LiLS sample from batch 2. The red curve is fitted with a complementary error function. The fitted Compton edge in this example is at 67.92 pC (p0).}
  \label{Cs137Spectrum_B2S1}
\end{figure}

Each sample was placed in a reflective PTFE collar on the face of a 1.5-inch (38.1 mm) Hamamatsu PMT (R9420-100) operated at about $5\times10^{5}$ gain (Figure~\ref{PSDsetup}). 
A Tektronix 3450 oscilloscope was used to record 20000 waveforms at 5 GHz sampling rate, with a 2 $\mu$s window and an 80-mV threshold. 
Another 5000 waveforms were acquired with an additional $^{137}$Cs source (0.68 MBq) to provide energy calibration. 
Figure~\ref{Cs137Spectrum_B2S1} shows the measured $^{137}$Cs spectrum for the sample from batch 2.

\begin{figure}[htb]
  \begin{center}
       \vspace{-0.3cm}
          \begin{minipage}[b]{0.5\textwidth}
                      \centering
            \includegraphics[width=7.5cm]{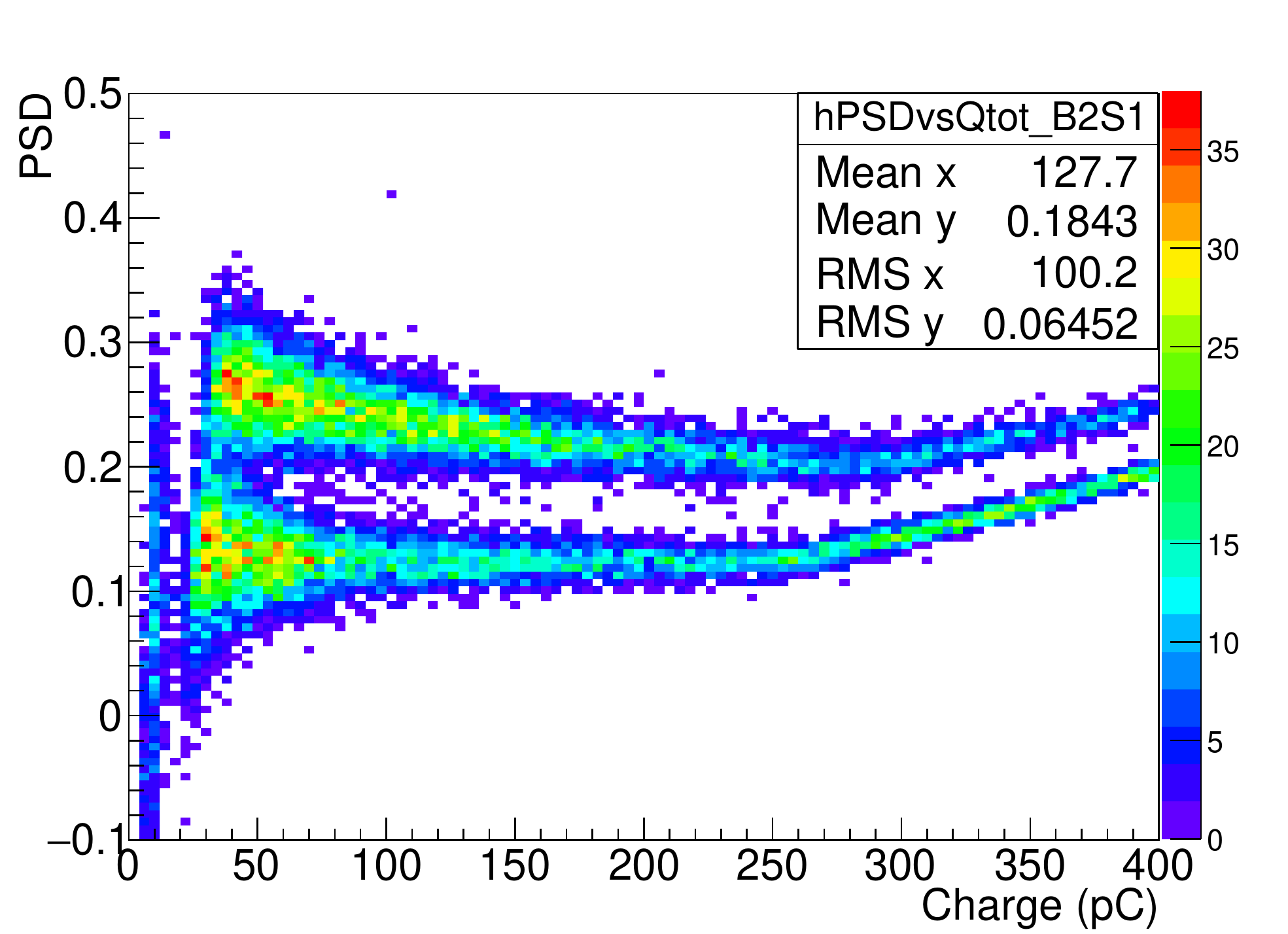}
          \end{minipage}
          \begin{minipage}[b]{0.5\textwidth}
            \centering
            \includegraphics[width=7.5cm, height=5.0cm]{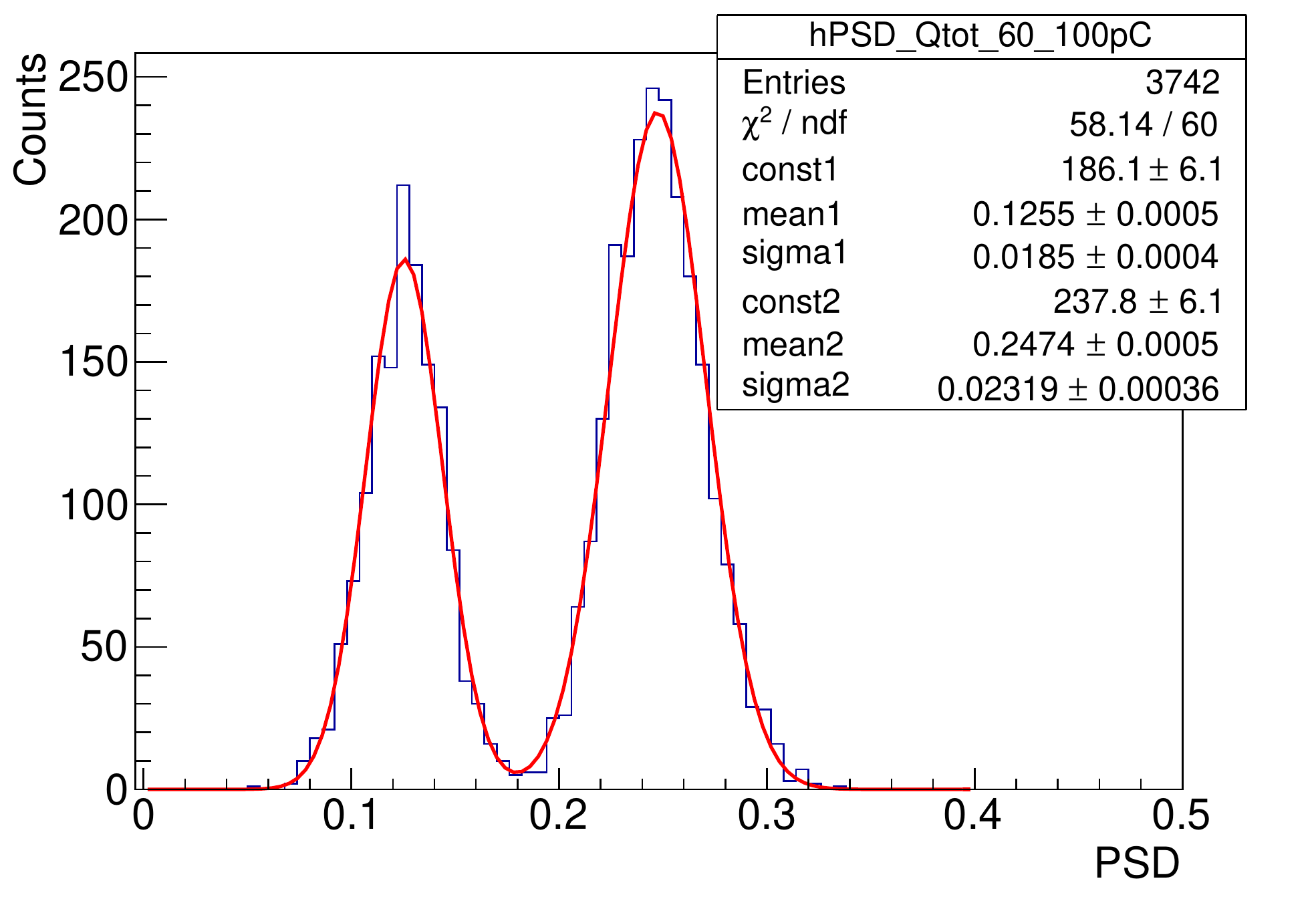}
          \end{minipage}%
        \vspace{-0.3cm}
        \caption{\textit{Left}: PSD versus total charge for the LiLS sample from batch 2. The upward trend at high 
        charge 
        is due to signal saturation during measurement. 
        \textit{Right}: The fitted PSD distribution for total charge between 60 and 100 pC. 
        For this sample a total charge of  77 pC corresponds to 540 keV visible energy. 
        }
        \label{PSD_B2S1_example}
  \end{center}
\end{figure}

The total charge, $q_{tot}$, measured by a waveform is computed from an interval of [-10, 160] ns about the minimum of the negative-going signal waveform. 
The Compton edge of the $^{137}$Cs data is fitted to estimate the energy scale. The PSD is defined as 
\begin{equation}
PSD=(q_{tot}-q_{f})/q_{tot},
\end{equation}
where $q_{f}$ is the charge in the [-10, 20] ns interval about the waveform minimum. The PSD distribution in intervals of charge is fitted with two Gaussians to evaluate the PSD for electronic and nuclear recoils (Figure~\ref{PSD_B2S1_example}).
The figure-of-merit (FOM) is defined as 
\begin{equation}
FOM=(\mu_{n}-\mu_{e})/\sqrt[]{FWHM_{n}^{2}+FWHM_{e}^{2}},
\end{equation}
\noindent where $\mu_{n(e)}$ and $FWHM_{n(e)}$ are the mean and full width at half maximum of the Gaussians corresponding to nuclear (electronic) recoils.

\begin{figure}[h]
  \begin{center}
       \vspace{-0.3cm}
        \hspace{-5cm}
          \begin{minipage}[b]{0.5\textwidth}
            \centering
            \includegraphics[width=16cm]{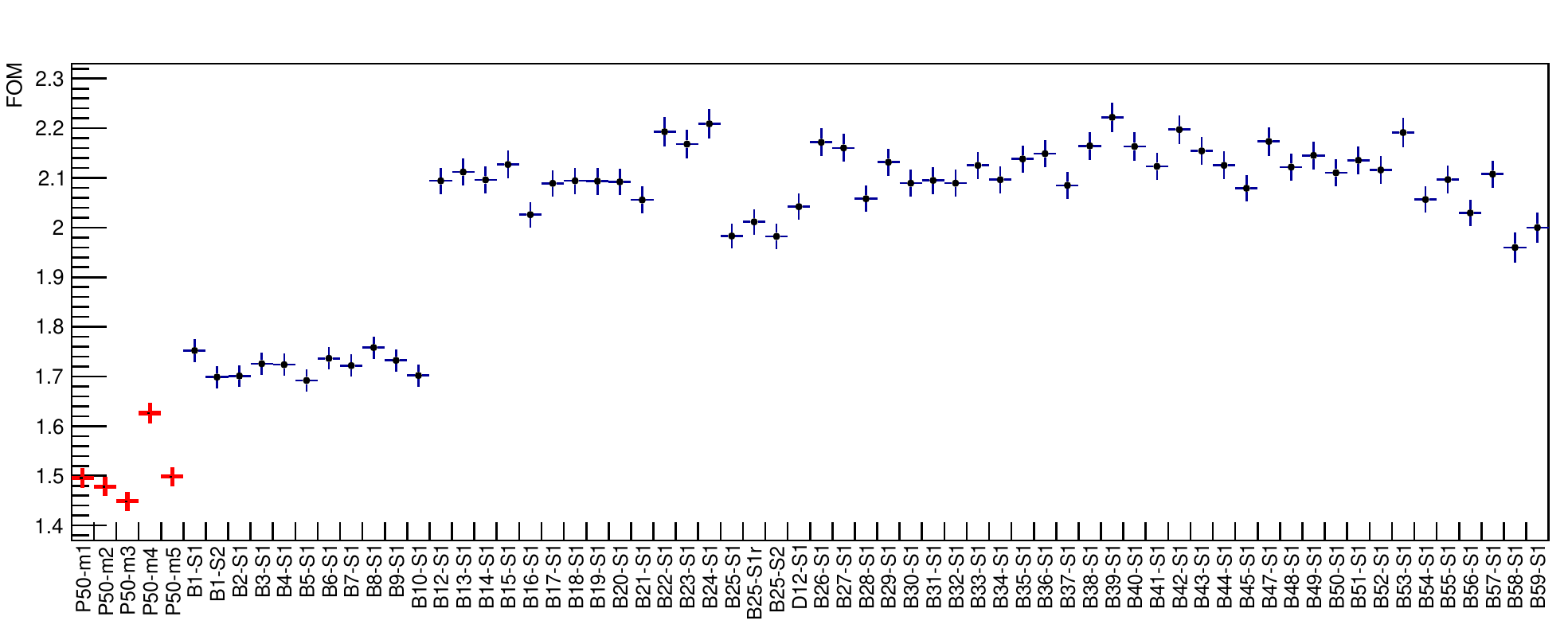}
          \vspace{-0.4cm}
          \end{minipage}\newline%
         \begin{minipage}[b]{0.5\textwidth}
            \centering
            \includegraphics[width=8cm]{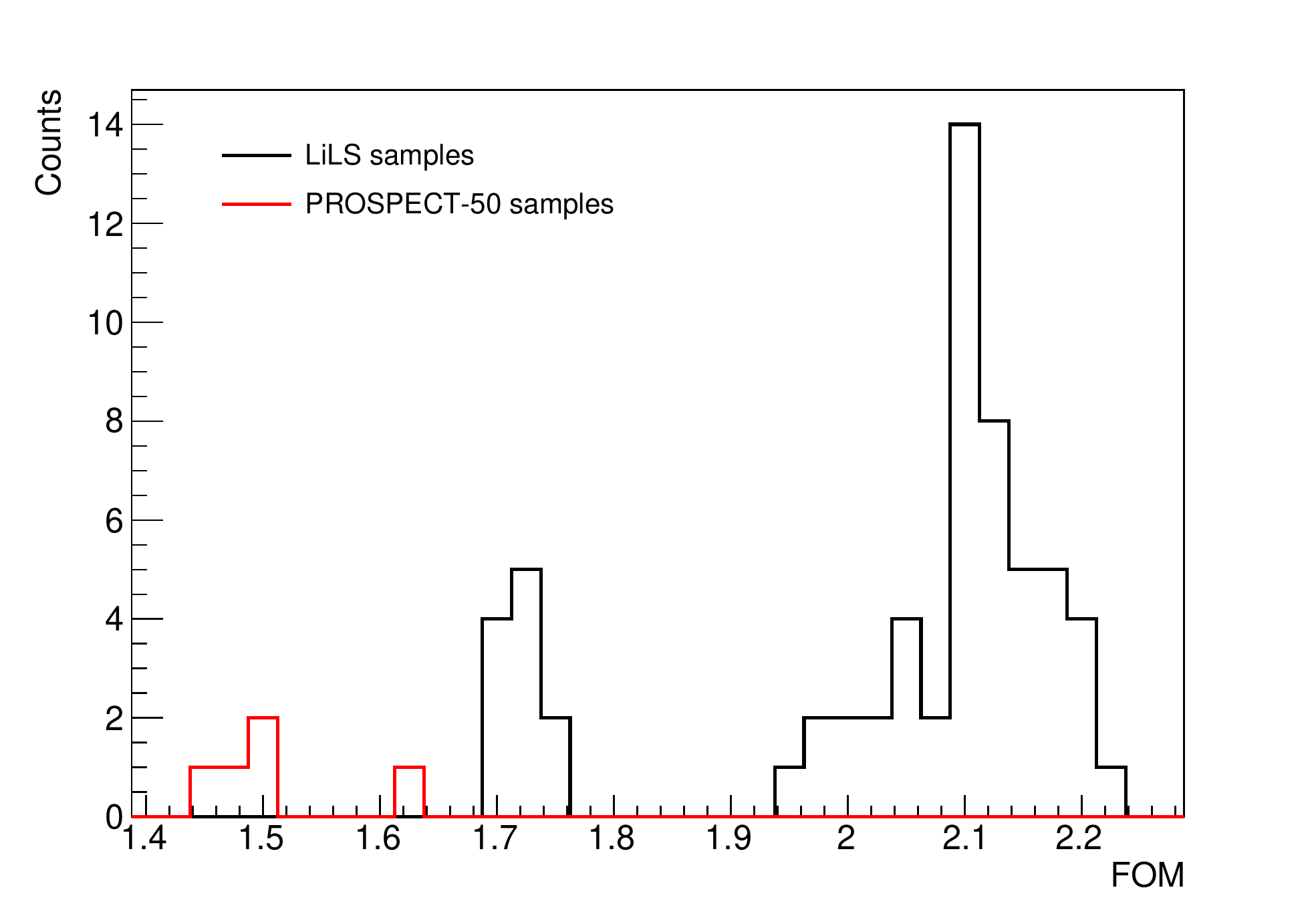}
          \end{minipage}%
        \vspace{-0.3cm}
        \caption{The measured FOM (n, $^{6}$Li) for the LiLS samples ($\it{top}$) and the distribution of the FOM values ($\it{bottom}$). The results for the PROSPECT-50 samples are also presented.}
        \label{PSD_FOM}
  \end{center}
\end{figure}

The measured FOM as a function of energy is linearly interpolated to estimate the FOM of the neutron capture process $^{6}$Li(n, $\alpha$)$^{3}$H at $\sim$540 keV electron equivalent which is used to assess the LiLS PSD quality. 
The higher the FOM value, the better the separation between nuclear and electronic recoils. 
The FOM of all measured samples exceeds that of the PROSPECT-50 samples as shown in Figure~\ref{PSD_FOM}, consequently the samples are all accepted. 
The increase in FOM for batch $\geq$12 is attributed to the better quality of EJ-309 scintillator from the later drums as described earlier.
\FloatBarrier

\section{Summary}

A total of fifty-nine batches of 90 liter LiLS were produced for the PROSPECT experiment. 
A one liter sample was collected from each batch for QA measurements.
Two batches were rejected due to unsatisfactory absorbance, another two batches were used for prototyping and material compatibility tests, the remaining batches satisfied the acceptance criteria in absorbance, light yield and PSD capabilities and were delivered for deployment in the PROSPECT detector at ORNL.


\acknowledgments

This material is based upon work supported by the following sources: U.S. Department of Energy (DOE) Office of Science, Office of High Energy Physics under Award No. DE-SC0016357 and DE-SC0017660 to Yale University, under Award No. DE-SC0017815 to Drexel University, under Award No. DE-SC0008347 to Illinois Institute of Technology, under Award No. DE-SC0016060 to Temple University, under Contract No. DE-SC0012704 to Brookhaven National Laboratory, and under Work Proposal Number SCW1504 to Lawrence Livermore National Laboratory. This work was performed under the auspices of the U.S. Department of Energy by Lawrence Livermore National Laboratory under Contract DE-AC52-07NA27344 and by Oak Ridge National Laboratory under Contract DE-AC05-00OR22725. This work was also supported by the Natural Sciences and Engineering Research Council of Canada (NSERC) Discovery program under grant \#RGPIN418579 and Province of Ontario.

Additional funding was provided by the Heising-Simons Foundation under Award No. \#2016- 117 to Yale University. J.G. is supported through the NSF Graduate Research Fellowship Program and A.C. performed work under appointment to the Nuclear Nonproliferation International Safeguards Fellowship Program sponsored by the National Nuclear Security Administration's Office of International Nuclear Safeguards (NA-241).

\end{document}